# Engineering Spin Splitting in Antiferromagnets by Superatoms with Internal Degree of Freedom

Fengxian Ma,[a,#] Zeying Zhang,[b,#] Zhen Gao,[a] Xiaobei Wan,[a] Yandong Ma,[c,*] Yalong Jiao,[a,*] Shengyuan A. Yang [d,*]

[a] College of Physics, Hebei Key Laboratory of Photophysics Research and Application, Hebei Normal University, Shijiazhuang 050024, China.
[b] College of Mathematics and Physics, Beijing University of Chemical Technology, Beijing 100029, China.
[c] School of Physics, State Key Laboratory of Crystal Materials, Shandong University, Jinan 250100, China
[d] Research Laboratory for Quantum Materials, Department of Applied Physics, The Hong Kong Polytechnic University, Kowloon, Hong Kong, China

*E-mail:

yandong.ma@sdu.edu.cn; yalong.jiao@hebtu.edu.cn; shengyuan.yang@polyu.edu.hk

[#]These authors contribute equally.

**ABSTRACT**

Superatoms, stable atomic clusters acting as building blocks for new materials, offer unique opportunities due to their rich properties and potential for 2D material assembly. While extensive research has focused on their similarities to ordinary atoms, the role of their internal degrees of freedom (IDOF) remains largely unexplored. Concurrently, compensated antiferromagnets (AFMs) with intrinsic spin-split band structures have emerged as a promising class of materials for spintronics, yet their experimental realization, particularly in two dimensions, is limited. Here, we bridge these two fields by proposing a novel strategy to achieve spin-split AFMs using superatoms with IDOFs. We establish our core concept using a simple model, demonstrating how superatom IDOFs can be leveraged to engineer system symmetry and induce spin splitting in AFM states. We concretely illustrate this strategy by first-principles calculations on a Mo-decorated carborophene sheets, constructed from *closo*-carborane superatoms. We show that the distinct IDOFs of carborane isomers (electric-dipole-like and nematic) are critical in determining the symmetry of the resulting 2D superatomic crystal and, consequently, the spin splitting pattern of its AFM states. Our findings underscore the profound significance of superatom IDOFs—a feature absent in ordinary atoms—and introduce a new paradigm for engineering spin splitting in AFM lattices. This work opens novel avenues for the design of advanced spintronic and quantum materials based on superatoms.

# I. INTRODUCTION

Superatoms refer to stable atomic clusters that can serve as basic building blocks for making a new class of cluster-assembled materials, just like atoms as building blocks for conventional materials [1,2]. The concept has been attracting tremendous interest, and a variety of superatoms have been realized and extensively studied in the past [3]. Recent advances have demonstrated the successful assembly of two-dimensional (2D) periodic lattices using the superatom $C_{60}$, paving the way for fabricating 2D materials from superatom units [4-6]. Superatoms may be composed of homo or heteroatomic species, with size ranging from a few to a few thousand atoms, hence they are expected to exhibit rich properties [7]. So far, research on superatoms has focused on topics such as their valence states [8], electronic orbitals [9], interaction with ligands [10], and stability rules [11], aiming to explore their similarity to elements in the periodic table [12]. However, unlike normal atoms, a superatom generally introduces a composite electronic environment that can interact nontrivially with the host lattice in the assembled material. Particularly, a superatom may possess certain additional *internal degrees of freedom* (IDOF), with no parallel in normal atoms. Such new IDOF can perturb the surrounding electronic and magnetic structures, affect symmetry and even topology of an extended system, and lead to novel emergent phenomena. This fascinating aspect of superatoms with IDOF has remained underexplored till now.

Meanwhile, a recent research focus in on compensated antiferromagnets (AFMs) with spin-split band structures [13]. Conventional AFMs are generally assumed to possess spin-degenerate energy bands. As a consequence, they are often considered incapable of supporting spin-dependent phenomena, such as spin current generation[14] and anomalous Hall effect [15]. This renders conventional AFMs, in many respects, similar to nonmagnetic materials—invisible to macroscopic electrical and optical probes—and thus limits their potential applicability in spintronic devices, in comparison to ferromagnets. Recently, it was realized that spin splitting is actually an intrinsic feature for a large class of AFMs [13,16-20]. The key is symmetry. Namely, the spin degeneracy in conventional AFMs is usually enforced by inversion ($I$) or some fractional translation that connects the two magnetic sublattices. If these symmetries are broken, then the bands should generally be spin-split. For collinear AFMs preserving certain combined symmetry of rotation and time reversal operations, the spin splitting pattern is alternating along different directions in the momentum space, and such systems are termed as altermagnets [21,22]. If such a combined symmetry (which connects the two sublattices) is also broken, then one may even have a half metal state with fully spin polarized bands in an energy window. These novel spin-split AFM states are generally known as AFMs with nonrelativistic spin splitting (to stress that the spin splitting is not from relativistic spin-orbit coupling) [23]. Due to their spin-split band structure, these AFMs may feature characteristics of ferromagnets, and a multitude of interesting properties have been discovered in them [24]. Despite the rapid progress, the experimentally realized spin-split AFMs [25], especially in 2D, are still very limited. In this context, developing a new strategy to generate spin-split AFMs, preferably with large spin splitting and fully spin polarized carriers, is of fundamental importance for unlocking the full potential of AFMs in spintronics.

Motivated by these advances and challenges, especially the recent experimental success in realizing 2D superatom-assembled crystals, in this work, we establish a link between the two previously disconnected fields, by proposing a general strategy to achieve spin-split AFMs using IDOF of superatoms. We illustrate our essential idea using a simple model, showing how IDOF of superatoms can be utilized to engineer the symmetry of the system, and to achieve spin splitting in AFM state. Then, we demonstrate the strategy by constructing a 2D hexagonal lattice composed of Mo and carborane cage clusters acting as superatoms. Carboranes—an important class of σ-electron delocalized boron–carbon clusters—are known for their exceptional chemical stability and have reached commercial availability [26]. Interestingly, carborane clusters have three different isomers, depending on the carbon positions in the cluster. This endows the carborane superatoms with a nontrivial IDOF. Specifically, the *closo*-1,2- and *closo*-1,3-carboranes have an electric-dipole like IDOF, whereas the *closo*-1,4-carborane has a nematic IDOF. We show that these IDOFs play a critical role in determining the symmetry of the 2D superatomic crystal as well as the spin splitting of AFM states. Our work unveils the significance of IDOFs of superatoms, a unique feature that is absent in ordinary atoms. It also highlights a new paradigm for engineering spin splitting in AFM lattices by incorporating superatoms, opening novel opportunities for spintronic and quantum materials design.

## II. METHODS

First-principles calculations were performed based on density functional theory (DFT), as implemented in the Vienna ab initio Simulation Package (VASP) [27,28]. The electronic exchange-correlation functional was treated using the generalized gradient approximation (GGA) proposed by Perdew, Burke, and Ernzerhof (PBE) [29]. To address the possible correlation effects from 4d orbitals of Mo, a Hubbard U correction with the typical value of U = 4.0 eV was applied [30]. For improved accuracy in band gap values, the more advanced hybrid Heyd-Scuseria-Ernzerhof (HSE06) functional [31] was employed. The plane wave energy cutoff was set to 500 eV. The structures were fully relaxed until the maximum force on each atom was below 0.005 eV/Å. The energy convergence criterion in the self-consistent calculations was set to $10^{-6}$ eV. A Gamma-centered Monkhorst-Pack k-point mesh, with a resolution of $2\pi \times 0.03$ Å$^{-1}$, was used for both geometry optimization and self-consistent calculations. To suppress artificial interactions between neighboring images of a 2D system, a vacuum slab of at least 10 Å in the z-direction was included. Phonon dispersion was computed using the Phonopy code [32] within the density functional perturbation theory [33]. Ab initio molecular dynamics (AIMD) simulations were conducted at 400 K to assess the thermal stability of the obtained 2D structures.

## III. RESULTS AND DISCUSSION

We first illustrate our essential idea using a simple AFM model. Consider a hexagonal lattice with AFM ordering, as shown in Fig. 1a. Here, red and blue colors denote sites with opposite local spins, and in the model, they correspond to local exchange field with opposite signs. Clearly, this model preserves the combined spacetime inversion symmetry $IT$, namely, the model is invariant under an inversion operation $I$ with respect to the center of a hexagon, followed by a time reversal operation $T$ which reverses the local spin direction (as indicated in Fig. 1a). Because of this $IT$ symmetry, the electronic band structure is guaranteed to be spin degenerate everywhere in momentum space, as schematically shown in Fig. 1d (and a model result is presented in Supplemental Material Fig. S1a).

One may try to induce spin splitting in band structure, e.g., by adding atoms at the center of the hexagons, as shown in Fig. 1b. However, putting a single atom at the central site cannot break the $IT$ symmetry, regardless of whether the atom carries nonzero magnetic moment or not (since magnetic moment is even under inversion $I$).

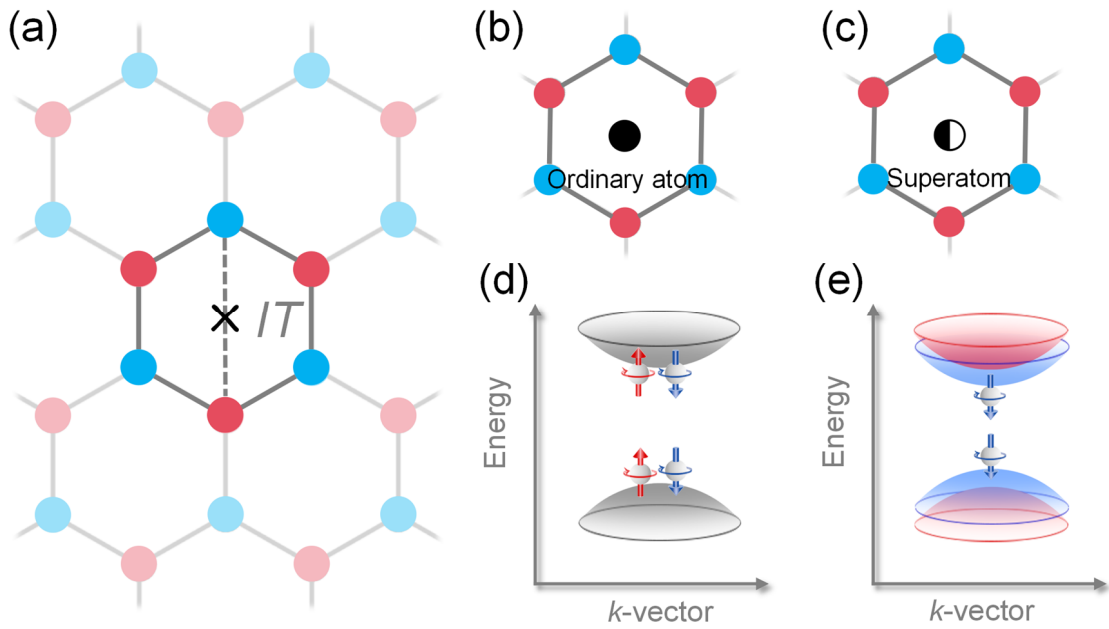

Figure 1. (a) A simple model of hexagonal lattice with AFM ordering that preserves $IT$ symmetry. Red and blue dots represent sites with opposite local spin moments. The cross indicates the inversion center. (b) Introducing an ordinary atom at the center of each hexagon cannot break the $IT$ symmetry. (c) Introducing a superatom with nonzero electric dipole IDOF breaks $IT$. (d) Band structure is spin degenerate under $IT$ symmetry, which is the case for (a) and (b). (e) Spin splitting is induced in AFM state for case (c) since $IT$ symmetry is broken.

This impossible mission for ordinary atoms can actually become possible by using superatoms with IDOF. For example, if the superatom could possess a nonzero electric dipole moment (as we demonstrate in concrete examples below), placing it at the center of the hexagon will naturally break the $IT$ symmetry (Fig. 1c) and hence generate spin splitting of the AFM band structure, as shown in Fig. 1e (and Supplemental Material Fig. S1b). From this simple model, one can see that superatoms can indeed differ from ordinary atoms in an essential way. Their unique IDOFs, which are not present in ordinary atoms, offer rich possibilities to engineer the symmetry and various material properties.

We demonstrate our strategy with a concrete material example, based on the carborane superatoms. Carboranes were first synthesized in the 1950s, although the results were not published until 1962-1963 [34,35]. The most common form, known as *closo*-carborane, consists of two carbon atoms and ten boron atoms arranged at the vertices of an icosahedron (see Fig. 2b-2e) [36,37]. *Closo*-carboranes have exceptional thermal and chemical stability, and are well-known as robust building blocks for constructing a wide range of functional materials [38,39]. However, they have not yet been employed in the construction of 2D crystals.

There are three stable isomers of *closo*-carborane $C_2B_{10}H_{12}$, depending on the relative position of the two carbon atoms in the icosahedral cage. In the structure, atoms are typically numbered starting from the apex atom with the fewest bonds, continuing clockwise within the hexagonal plane, with the carbon atoms assigned the lowest possible numbers (Fig. 2b). As a result, the three *closo*-carborane isomers may be labeled as: 1,2- (ortho-), 1,3- (meta-), and 1,4- (para-) carborane, as shown in Fig. 2b-2d. The three isomers can be selectively synthesized [35,40,41]. The ortho-isomer is typically produced by reacting decaborane ($B_{10}H_{14}$) with acetylenes in the presence of a Lewis base. The meta- and para-isomers are generated via irreversible thermal isomerization of the ortho-isomer, occurring at 465-500 °C and 615-700 °C, respectively [42]. Today, all three isomers are commercially available. It is also noted that *closo*-carboranes are isoelectronic with *closo*-boranes such as icosahedral $B_{12}$ cage (Fig. 2e), which makes them structurally comparable. In the following, we shall explore how *closo*-carborane can form a stable 2D crystal, and we also include icosahedral borane for comparison.

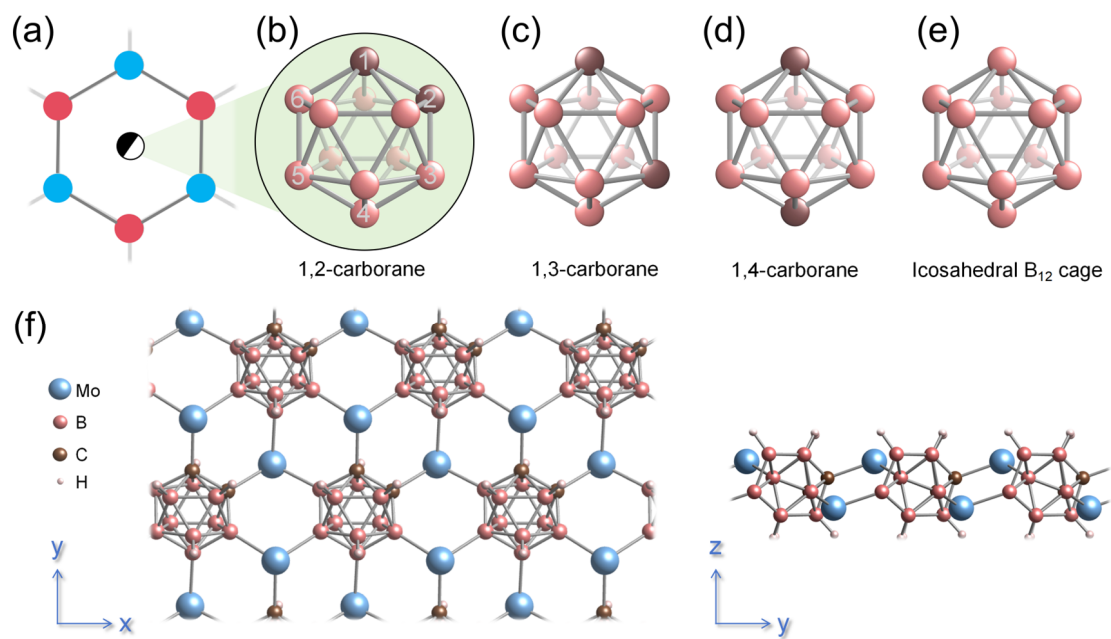

Figure 2 (a–b) Schematic illustration for the use of 1,2-carborane as a superatom to occupy the central site of a hexagonal lattice. (b-e) Structural representations of 1,2-, 1,3- and 1,4-carboranes, and also borane ($B_{12}$ cage). The surrounding hydrogen atoms are not presented in the figure here for simplicity. (f) Top and side views of the 1,2-carboraphene sheet. The structures of 2D sheets with other superatoms are similar.

Isolated *closo*-carboranes exhibit aromaticity and multi-center two-electron bonding, which compensate for electron deficiency and ensure overall cluster electronic stability. To construct a stable 2D hexagonal lattice, terminal hydrogen atoms on $C_2B_{10}H_{12}$ must be removed to create reactive open sites on the cage vertices. These vacant positions expose boron or carbon atoms with unsaturated valencies, enabling them to act as electron acceptors and facilitate direct orbital overlap with foreign atoms.

In our design, we select molybdenum (Mo) as the foreign atom. As a transition metal with half-filled 4$d$ orbitals (electron configuration: 4$d^5$5$s^1$), Mo can form directional covalent bonds via effective overlap between its *d* orbitals and the *p* orbitals of the carborane clusters. Additionally, the 5*s* orbital promotes electron delocalization across the lattice. Mo's relatively high *d*-orbital energy and moderate electronegativity provide a balance between electron donation and back-donation, stabilizing the metal–boron/carbon bonds without inducing excessive charge transfer that could disrupt the cluster's structural integrity. In addition, when forming a 2D hexagonal lattice with the carborane superatoms, Mo tends to have valence +1 and its partially filled *4d* shell will have a nonzero magnetic moment, leading to magnetism, which is the desired feature we are looking for.

**Table 1** The lattice constants (in Å), layer thickness (*d*, in Å), cohesive energy $E_{co}$ (in eV/atom), magnetic ground state (MGS), magnetic easy axis, HSE band gaps (in eV), spin splitting and Néel temperature $T_N$ for the designed 2D superatom sheets.

| System | Lattice constants | d | $E_{co}$ | MGS | Easy axis | Band gaps | spin splitting | $T_N$ |
|---|---|---|---|---|---|---|---|---|
| 1,2-carboraphene | a=6.47, b=6.48 | 4.62 | -5.27 | nAFM | z | 3.11 | Yes | 395 K |
| 1,3-carboraphene | a=6.42, b=6.49 | 4.57 | -5.29 | nAFM | z | 2.68 | Yes | 357 K |
| 1,4-carboraphene | a=b=6.45 | 4.54 | -5.30 | nAFM | z | 3.34 | No | 341 K |
| $B_{12}$ monolayer | a=b=5.71 | 4.67 | -5.06 | nAFM | x | 1.89 | No | 280 K |

The resulting stable 2D Mo–carborane networks are shown in Fig. 2f (using 1,2-carboraphene as a representative example). We term these 2D materials "carboraphene", as they represent the first constructed carborane-cage-based 2D crystals. The structure adopts a hexagonal lattice, with each carborane superatom hexacoordinated to Mo atoms. We confirm the dynamical, thermodynamic, and mechanical stabilities of all three 2D Mo–carborane isomers as well as 2D Mo-borane (with *closo*-carborane replaced by *closo*-borane in the structure) through phonon spectrum analysis (Fig. S2), *ab* initio molecular dynamics (AIMD) simulations (Fig. S3), and compliance with the Born stability criteria (details are presented in Table S1), indicating their potential for experimental synthesis. Particularly, the AIMD simulations show that 2D Mo–carboranes can maintain a robust structure up to 400 K, demonstrating an excellent stability.

As mentioned, Mo carries nonzero magnetic moments, so we investigate the magnetic ground state configuration for the four 2D superatom crystals. Four typical magnetic configurations are considered, including one ferromagnetic and three AFM ones, as illustrated in Fig. 3a-3d. Our first-principles calculations indicate that the Néel-type AFM ordering (nAFM) is the ground state for all the four materials (Table S2). The magnetic moments are mainly on the Mo site, with a value of $\sim 3\mu_B$ per Mo.

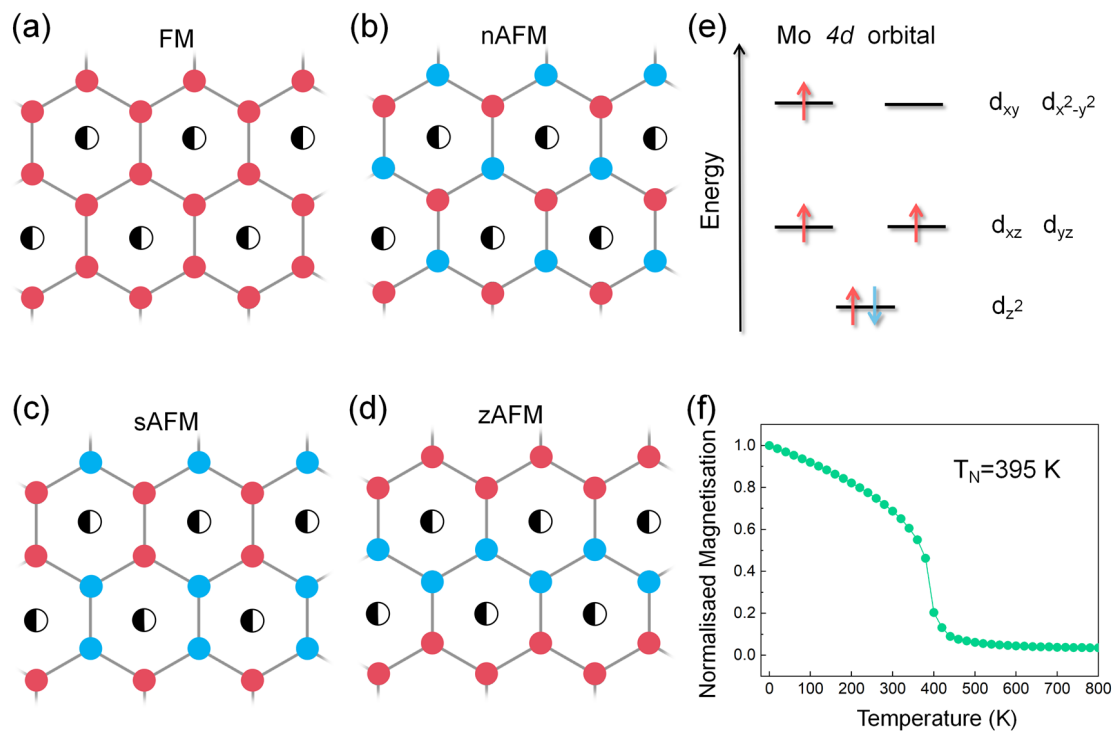

Figure 3 (a-d) Typical FM and AFM configurations considered in our calculation. Red and blue dots represents sites with opposite local spin directions. (e) Splitting pattern of the Mo 4*d*-orbitals. (f) Normalized magnetic moment of 1,2-carboraphene as a function of temperature, obtained from Monte Carlo simulations.

The value of magnetic moment is consistent with the valence state of Mo. In these materials, Mo has a valence of +1, whereas the superatoms have a valence of -2. Indeed, charge analysis indicates ~1 electron transfer from Mo to the cage clusters—0.90 e for 1,2-carboraphene, 1.06 e for 1,3-carboraphene, 0.87 e for 1,4-carboraphene, and 1.03 e for the $B_{12}$ cage sheet. (This sizable charge transfer also confirms Mo is an effective donor, which compensates for electron deficiency and thereby stabilizes the structures.) It follows that the Mo ion has a $d^5$ electron configuration. In the trigonal planar geometry, the five $d$ orbitals split into three group, as illustrated in Fig. 3e, with $d_{z^2}$ orbital being the lowest. Since trigonal planar geometry generally results in weak ligand field splitting, a high-spin configuration is expected. This leads to three unpaired spins, consistent with the calculated magnetic moment of $\sim 3\mu_B$ per Mo.

To investigate the robustness of the AFM ordering, we perform Monte Carlo simulations. The result shows that the Néel temperature $T_N$ of 1,2-carboraphene is the highest, reaching 395 K (see Fig. 3f), which is substantially higher than many well-known 2D magnets discovered so far, such as $CrI_3$ and $Cr_2Ge_2Te_6$ [43,44]. $T_N$ values for 1,3- and 1,4-carboraphene are found to be 357 K and 341 K, respectively (see Fig. S4), which are also above room temperature. Such room-temperature 2D magnetism is very much desired for applications.

Next, we study the band structures of these 2D superatom AFMs, and the focus is on the possible spin splitting in the AFM state. In Fig. 4, we present the calculated band structures for the four 2D AFMs on the PBE level without spin-orbit coupling. (We have checked that including spin-orbit coupling only has weak effect on the low-energy band structure.) And the HSE results are presented in Fig. S5, which do not show qualitative differences. One can see that all four materials are AFM insulators. The three carboraphenes have gap values around 2.12 to 2.68 eV, and the value for the $B_{12}$ cage sheet is ~0.69 eV. The corresponding HSE gap values are also shown in Table 1.

Importantly, one observes that the four materials exhibit distinct features in terms of spin splitting. The 1,2- and 1,3-carboraphene show sizable spin splitting in the band structure, whereas the 1,4-carboraphene and the $B_{12}$ cage sheet have no spin splitting at all. In this sense, 1,4-carboraphene and $B_{12}$ cage sheet are like conventional AFMs, while 1,2- and 1,3-carboraphene represent the spin-split AFMs. It is worth noting that the spin-split AFMs here are not of altermagnetic type: The spin splitting is not alternating in momentum space, instead, one can have fully spin polarized bands in an energy window, which is especially evident in Fig. 4b. This will be an important advantage in generating highly spin-polarized current.

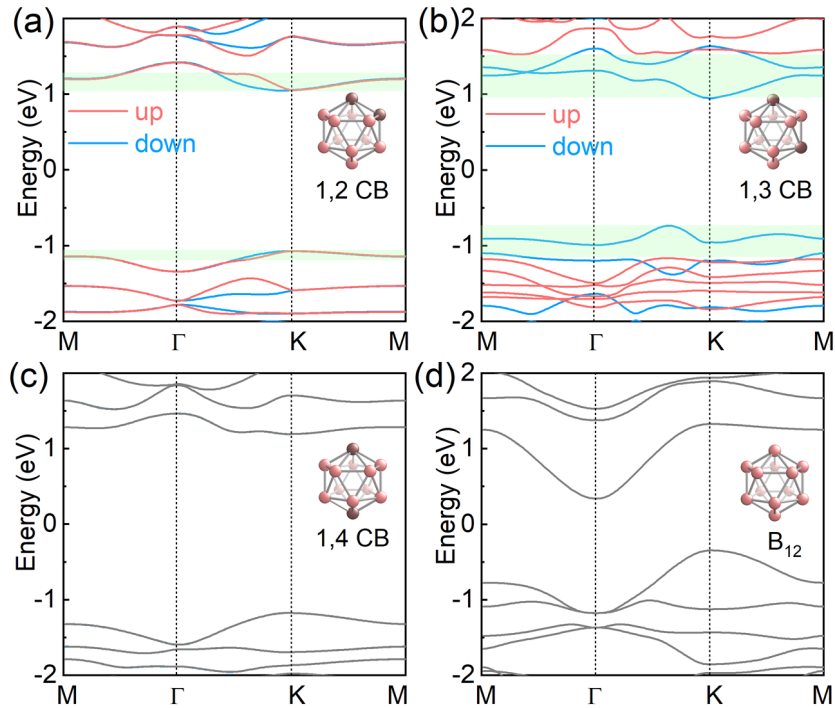

Figure 4 (a-d) Band structures of 1,2-, 1,3-, and 1,4-carboraphene (CB), along with the $B_{12}$ cage-based 2D network. Green regions highlight the spin splitting around band edges in 1,2- and 1,3-carboraphene. In (c) and (d), the bands are fully spin degenerate.

How to understand the reason behind this big difference? Here, we can go back to the simple model in Fig. 1. Without the superatoms, the Mo atoms just form an AFM honeycomb lattice, similar to that in Fig. 1a. Hence, the spin splitting we see in Fig. 4 is associated with placing the superatoms in the Mo hexagons, like that in Fig. 1c. Now, although the four superatoms look quite similar, they make qualitative differences in the spin splitting, because they have different IDOF. For isolated 1,2- and 1,3-carborane cages, they are clearly asymmetric, with a polar point group $C_{2v}$. As a result, they both carry nonzero electric dipole moments as an IDOF (Fig. 5a and 5b). Specifically, due to the higher electronegativity of carbon compared to boron, adjacent carbon atoms in the 1,2-isomer create a localized region of high electron density (Fig. 5a), resulting in a pronounced dipole moment of approximately 3 Debye. Meanwhile, the 1,3-isomer exhibits a smaller dipole moment of ~1.5 Debye due to a more dispersed charge distribution. Because of this nontrivial IDOF associated with 1,2- and 1,3-carborane superatoms, placing them into the Mo lattice breaks the *IT* symmetry, and therefore generates the spin splitting. In contrast, the $B_{12}$ cage has a high symmetry with $I_h$ point group, which forbids an electric dipole IDOF (Fig. 5d), so it cannot lift the spin degeneracy. As for 1,4-carborane superatom, its symmetry is lower than the $B_{12}$ cage, possessing the $D_{5d}$ symmetry. This superatom is nonpolar, with a vanishing electric dipole, which can be easily understood from the fact that the two carbon atoms are sitting at opposite vertices of the icosahedral cage (Fig. 5c). Hence, it also cannot break the *IT* symmetry either. It is worth noting that although 1,4-carborane does not have an electric dipole IDOF, it has a nematic type IDOF, namely, the lines connecting the two carbon sites defines a nematic director for this superatom (Fig. 5c). This distinct type of IDOF may have important implications in other contexts.

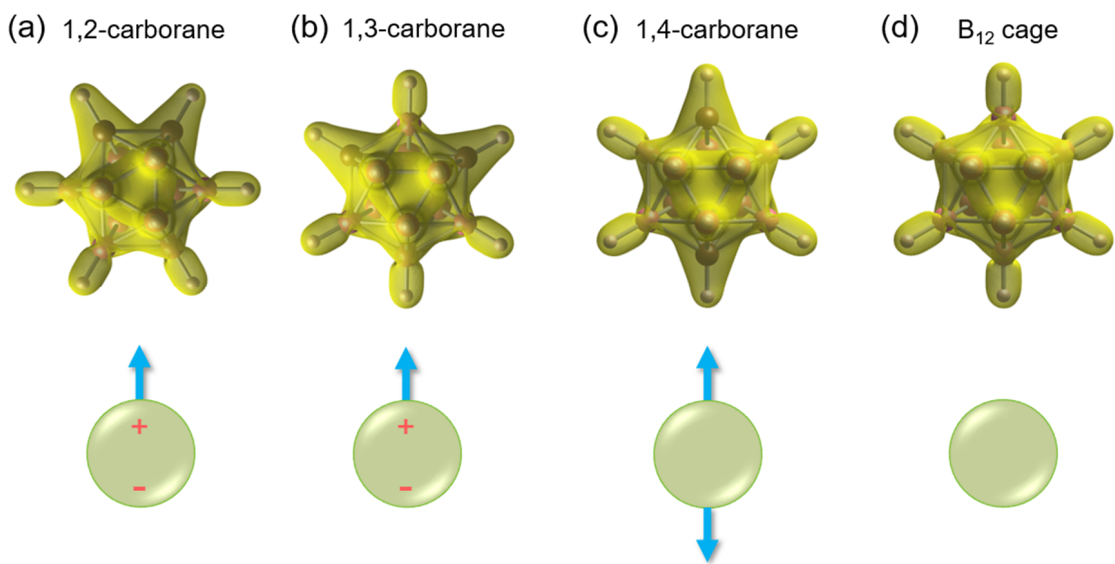

Figure 5 (a-d) The top panel shows the charge density distribution (isovalue of 0.1 e/Å

$^3$) of each superatom. The bottom panel schematic shows the IDOF of each superatom: 1,2- and 1,3-carboranes possess electric dipole IDOF, 1.4-carborone has a nematic type IDOF, and $B_{12}$ cage does not have these IDOFs.

We have demonstrated that the IDOF of superatoms can have significant impact on the spin splitting of the AFM band structure. It strongly affects other properties as well. For example, the polar carborane cage generates net attractive forces with neighboring Mo atoms, affecting the in-plane stiffness. Our calculation indicates that the Young's modulus along the *x*-direction for 1,2-, 1,3-, and 1,4-carboraphene is 35.88, 34.99, and 24.88 N/m, respectively, clearly demonstrating that the cage polarity controls the mechanical strength (Fig. S6). Additionally, the electric dipole influences the local electrostatic environment of the crystal field, which can affect the magnetic anisotropy energy (Fig. S7-S8). Beyond these effects, tuning the dipole moment of superatoms may enable modulation of optoelectrical[45], dielectric[46], catalytic[47], and electrostatic potential[48] properties, as well as structure arrangement[49]. These will be interesting topics to explore in future works.

We have a few remarks before ending. First, the key message we want to convey in this work is that IDOF of superatoms endows them with much richer possibility beyond ordinary atoms. Here, we showed how this IDOF can be utilized to generate large spin splitting in compensated AFMs, a hot topic of current research. This is just one simple application. As mentioned above, by altering symmetry characters and chemical environment, IDOF of superatoms can impact various physical and chemical properties of a superatom crystal.

Second, in the example discussed here, the key IDOF to break symmetry is the electric dipole of superatoms. Clearly, there are many other types of IDOF. The nematic type IDOF of 1,4-carborane is one example. This may lead to interesting consequences, e.g., in the kind of topological defects. Similar to the discussion in liquid crystals, the superatoms with nematic IDOF may form a half vortex or $\pi$-wall in a 2D lattice, if the nematic directors are confined in plane; and they may form point defect with $Z_2$ topological charge, if the directors are allowed to rotate out of plane. These features are distinct from those for the electric dipole IDOF.

Third, we used Mo-2s as examples to demonstrate our general idea. Nevertheless, due to their excellent properties, these examples themselves are quite interesting. We have demonstrated that as 2D materials, they have excellent stability, robust room-temperature magnetism, and large spin-splitting in AFM state. Particularly, for the 1,3-carboraphene, both conduction and valence band edges are fully spin polarized, capable of generating current with high spin polarization, just like a half-metal. All these features are desired for spintronic applications.

Finally, we presented carboranes as an example of superatoms capable of inducing symmetry breaking and unconventional magnetism in 2D materials, they are by no means unique. There already exist a rich family of superatoms with rich IDOFs, and there are countless possibility in constructing new superatoms. The incorporation of superatoms into host lattices (not confined to the 2D lattices) opens a broad avenue for symmetry engineering, potentially leading to a variety of emergent phenomena such as topological states of matter, nontrivial spin textures, Mott-like transitions, or flat-band magnetism. This offers a new design principle for functional quantum materials.

## IV. CONCLUSIONS

In summary, we propose a new strategy to induce spin splitting in AFMs by using superatoms, highlighting the role of IDOF of superatoms which is absent in ordinary atoms. Using carborane superatoms as an example, we demonstrate that the different isomers possess different electric dipole IDOF, and can effectively lift or preserve spin degeneracy in the electronic band structure. The resulting 2D Mo-carboraphene superatom crystals are room-temperature spin-split AFMs with excellent properties. Our work offers a general strategy for designing AFM materials with engineered spin polarization, opening new opportunities for low-power and field-free spintronic applications. Furthermore, this superatom approach is not just limited to the effect of spin splitting, but offers a general pathway to engineer material properties based on supseratoms with IDOFs.

## ACKNOWLEDGMENTS

The authors thank D. L. Deng for valuable discussions. This work is supported by the National Natural Science Foundation of China (Grant No. 12204144, and 11904077), Hebei Natural Science Foundation (Grant No. A2024205029, A2022205027 and A2021205024), and HK PolyU start-up fund (P0057929).